\documentclass[final,5p,times,twocolumn,sort,compress]{elsarticle}

\usepackage{amssymb} 
\usepackage{amsmath} 
\usepackage{lineno} 
\usepackage{tabularx}
\usepackage{booktabs}               	
\usepackage{multirow}               	
\usepackage{nicefrac}
\usepackage[table,dvipsnames]{xcolor}
\usepackage[separate-uncertainty=true,multi-part-units=single]{siunitx}
\DeclareSIUnit\mmHg{mmHg}
\DeclareSIUnit\cmHO{cmH2O}
\DeclareSIUnit\Fr{Fr}
\DeclareSIUnit\breaths{breaths}
\DeclareSIUnit\bpm{BPM}
\DeclareSIUnit\years{years}
\DeclareSIUnit\month{month}
\DeclareSIUnit\days{days}
\DeclareSIUnit\hours{hours}

\journal{Biomedical Signal Processing and Control}
\begin{document}
\begin{frontmatter}
\title{A quantitative analysis of intraventricular bioimpedance in an in~vivo pilot study \mbox{with contextual pressure measurements}}

\author[label1]{Fabian Flürenbrock\corref{cor1}}
\cortext[cor1]{Corresponding author: \texttt{ffluerenb@ethz.ch}}
\author[label2]{Christian T. Stoeck}
\author[label3]{Markus F. Oertel}
\author[label2]{Miriam Weisskopf}
\author[label1]{Melanie N. Zeilinger}
\author[label1]{\mbox{Marianne Schmid Daners}}
\author[label1]{Leonie Korn}
\affiliation[label1]{
    organization={Institute for Dynamic Systems and Control}, 
    addressline={Department of Mechanical and Process Engineering}, 
    city={ETH Zurich},
    country={Switzerland}}
\affiliation[label2]{
    organization={Center for Preclinical Development},
    city={University Hospital Zurich and University of Zurich},
    country={Switzerland}}
\affiliation[label3]{
    organization={Department of Neurosurgery},
    city={University Hospital Zurich},
    country={Switzerland}}

\begin{abstract}
Hydrocephalus is a neurological condition characterized by disturbed cerebrospinal fluid (CSF) dynamics and is typically treated with shunt systems that drain excessive CSF out of the ventricular system.
Continuous monitoring of ventricular CSF volume, however, remains a major unmet need in the clinical management of this condition. 
While intraventricular bioimpedance (BI) has been proposed as a potential marker of CSF volume, prior investigations have been limited to simulations, in~vitro phantoms, and small animal models.
This work presents the development of a measurement system for intraventricular BI and its evaluation in a large animal model. 
The measurement system was first validated in~vitro using a mechatronic test bench replicating physiological CSF dynamics and subsequently applied in an in~vivo pilot study with concurrent CSF and blood pressure monitoring.
Time series analysis of the recorded signals revealed physiological BI waveform components linked to the cardiac and respiratory cycles. 
In addition, changes in BI following CSF volume alterations induced through intrathecal bolus infusions of artificial CSF were observed and found to be correlated to changes in CSF and blood pressures.
These results provide the first in~vivo evidence in a large animal model that BI reflects CSF dynamics as well as cerebral hemodynamics. 
Complementing intracranial pressure and CSF drainage measurements in smart shunt systems with BI could enable more comprehensive patient monitoring and physiologically informed control of hydrocephalus therapy.

\end{abstract}
\end{frontmatter}

\section{Introduction}\label{sec:Introduction}
Hydrocephalus is a neurological condition characterized by disturbed cerebrospinal fluid (CSF) dynamics and an excessive accumulation of CSF within the cerebral ventricular system~\cite{Rekate2008}.
Without appropriate therapy, such as the implantation of a shunt system that drains the excessive CSF from the cerebral ventricular system into another body compartment where it can be absorbed, alterations in the intracranial pressure (ICP) can lead to severe neurological damage~\cite{Rigamonti2014}.
While ultimately the volume of CSF is to be regulated, contemporary shunts are based on passive mechanical valves for which the CSF drainage rate is determined by the pressure difference acting on the shunt valve~\cite{Symss2015}.
The various pressures on which this rate-determining pressure difference depends are not directly physiologically related and can be disturbed by external factors such as the patient's posture or activity~\cite{Gehlen2017, Browd2006b}.
Therefore, new mechatronic shunt systems that can actively alter the CSF drainage rate are being developed to improve hydrocephalus therapy \cite{Misgeld2015, Fluerenbrock2025FBCNS}.
However, to achieve the aspired goal of closed-loop CSF volume regulation, new sensor technologies for monitoring ventricular CSF volume must be developed first.
Current approaches to do so rely on medical imaging~\cite{Lundervold2000, Hladky2024}, which is costly and cannot be performed continuously, making them unsuitable for real-time patient monitoring and closed-loop shunt control.

\begin{figure*}[!t]
    \centering
    \includegraphics[width=\textwidth]{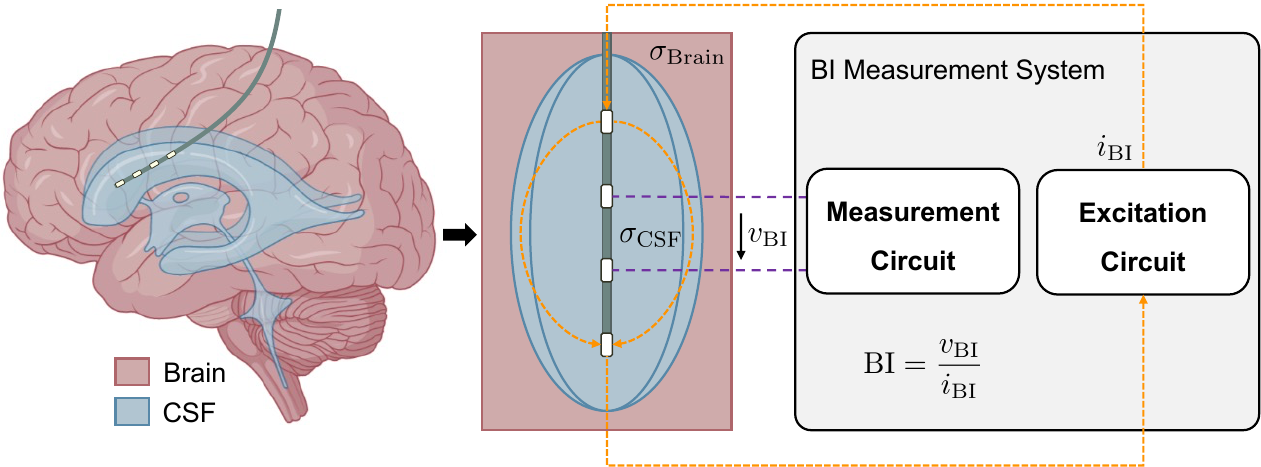}
    \caption{Sketch of the tetrapolar bioimpedance (BI) measurement principle based on an approximation of the complex cerebral ventricular geometry. An excitation current $i_\text{BI}$ is injected through the two outer electrodes, inducing a voltage drop $v_\text{BI}$ across the two inner electrodes. Both signals are acquired by the measurement system to compute the BI signal.}
    \label{fig:measurement_system}
\end{figure*}

Inspired by earlier work on conductance catheters used for the assessment of ventricular heart volume~\cite{Baan1981, Baan1984, Salo1986, Wei2005}, the measurement of intraventricular bioimpedance has been proposed as a tool to address the problem of CSF volume monitoring in hydrocephalus~\cite{Linninger2009a}.
Since the electrical conductivity of CSF is about one order of magnitude higher than the electrical conductivity of brain tissue (see Table~\ref{tab:electrical_conductivity}), any ventricular enlargement due to an increase in CSF volume should lead to a decrease in the intraventricularly measured bioimpedance (see Figure~\ref{fig:measurement_system}).
This hypothesis was confirmed in~silico with electromagnetic field simulations~\cite{Linninger2009a, Castelar2019}, in~vitro with ventricular brain phantoms made from agarose, silicone, or silicone-carbon~\cite{Linninger2009a, Basati2011, Basati2012, Castelar2019}, and in~vivo with hydrocephalic rats~\cite{Basati2011, Basati2012}.

To advance toward continuous monitoring of ventricular CSF volume, this work presents the investigation of intraventricular bioimpedance as a potential CSF volume marker in a large animal model.
To this end, a bioimpedance measurement system is built using commercially available impedance catheters and integrated measurement chips.
The developed system is first validated in~vitro using a mechatronic test bench capable of replicating physiological CSF dynamics.
Subsequently, the system is applied in an in~vivo pilot study based on a large animal model with contextual CSF and blood pressure measurements.

\begin{table}[t]
    \centering
    \setlength\extrarowheight{2pt}
    \caption{Electrical conductivity of biological tissues at \qty{50}{\kHz} \cite{Hasgall2022}}
    \label{tab:electrical_conductivity} 
    \small
    \begin{tabularx}{\columnwidth}{XX}
    \toprule
    \textbf{Tissue} & \mbox{\textbf{Electrical Conductivity [S/m]}} \\
    \midrule
    Blood                   & 0.7008 \\
    Brain (Grey Matter)     & 0.1275 \\
    Brain (White Matter)    & 0.0776 \\
    Cerebrospinal Fluid     & 2.0000 \\
    \bottomrule
    \end{tabularx}
\end{table}

\section{Methods}\label{sec:Methods}
This section provides an overview of the methods employed in this work.
In Section~\ref{sec:Methods_InVitro}, the design and in~vitro validation of the bioimpedance measurement system is described.
In Section~\ref{sec:Methods_InVivo}, the application of the measurement system in an in~vivo pilot study is reported and a quantitative analysis of the acquired data provided.

\subsection{Bioimpedance Measurement System}\label{sec:Methods_InVitro}
\subsubsection{System Design}\label{sec:Methods_InVitro_Design}
The bioimpedance measurement system developed in this work is based on an integrated bioimpedance measurement chip (AD5940, Analog Devices, Wilmington, MA, United States).
It is configured to perform a tetrapolar bioimpedance measurement, as sketched in Figure~\ref{fig:measurement_system}. 
In the excitation circuit, a zero-mean high-frequency alternating current is injected through the two dedicated excitation electrodes. 
To drive this current, a \qty{50}{\kHz} sinusoidal voltage signal generated by a digital-to-analog converter is applied to a transimpedance amplifier.
In the measurement circuit, the system records both the injected excitation current $i_\text{BI}$ through the transimpedance resistor and the voltage drop $v_\text{BI}$ across the two dedicated measurement electrodes that is induced by the excitation current. 
These signals are acquired using a single analog-to-digital converter, which multiplexes between the voltage and current measurement paths. 
The onboard microcontroller extracts the magnitude and phase of both measurement signals at the programmed excitation frequency by applying a discrete Fourier transform.
Finally, the bioimpedance signal is computed as $\text{BI} = \nicefrac{v_\text{BI}}{i_\text{BI}}$.

Since both the excitation current and the resulting voltage are measured, the system maintains phase accuracy and achieves impedance measurements that are robust to variations in the electrical characteristics of the measured object or material.
To comply with medical safety standards~\cite{IEC60601}, the root mean square (RMS) value of the injected current is kept below \qty{1}{\mA} through an appropriate scaling of the excitation voltage and a corresponding selection of the transimpedance resistor.
In addition to the bioimpedance subsystem, the developed measurement system also integrates a pressure-sensing subsystem based on a precision amplifier (OPA197IDR, Texas Instruments, Dallas, TX, United States). 
For simultaneous measurements of bioimpedance and pressure with a single catheter, a biomedical research measurement catheter (SPR-877, Millar, Houston, TX, United States) is used.
This \qty{3}{\Fr} ($\approx$ \qty{1}{\mm} outer diameter) measurement catheter features ten circumferential ring electrodes spaced over \qty{25}{\mm}, which can be combined into different tetrapolar (4-point) configurations to adapt the measurement setup to various anatomical or experimental geometries. 
Additionally, a pressure sensor based on a piezoresistive Wheatstone bridge is integrated within the electrode array for co-localized acquisition of ICP and BI signals.

\begin{figure*}[!t]
    \centering
    \includegraphics[width=\textwidth]{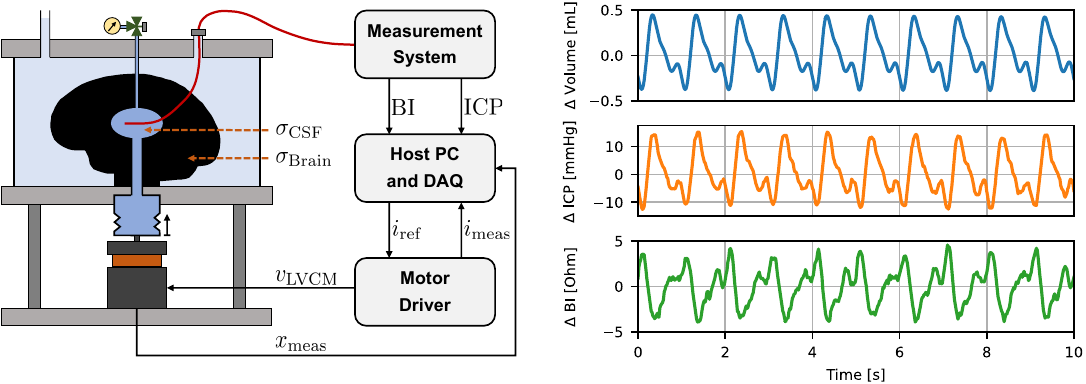}
    \caption{In~vitro validation of the developed measurement system. Left: Sketch of the experimental test setup~\cite{Castelar2022}. The intracranial pressure (ICP) and bioimpedance (BI) measurement catheter (shown in red) is inserted in the ventricle of the silicone-carbon brain phantom (shown in black), which is filled with artificial cerebrospinal fluid (CSF). Ventricular volume changes are induced by compressing the connected bellows using a linear voice coil motor (LVCM). Right: Signal sequences resulting from in~vitro testing with the mechatronic test bench. Shown are changes in ventricular CSF volume (top panel), ICP (middle panel), and BI (bottom panel).}
    \label{fig:invitro}
\end{figure*}

\subsubsection{System Validation}\label{sec:Methods_InVitro_Validation}
For experimental in~vitro validation of the introduced bioimpedance measurement system, a custom-built mechatronic test bench from~\cite{Castelar2022} is used that replicates physiological conditions such as ventricular volume changes and ICP waveforms.
The central element of this test bench consists of a silicone-carbon brain phantom, where carbon was added into the otherwise non-conducting silicone in order to mimic both the electrical and mechanical properties of the human brain ($\sigma_\text{Brain} \approx \qty{0.17}{\siemens\per\meter}$).
It is mounted within a cylindrical water tank and connected to a bellows, which in turn is attached to a linear voice coil motor (LVCM; CAL75, SMAC Corporation, Carlsbad, CA, United States). 
The ventricle of the brain phantom is an ellipsoid with a nominal volume of \qty{40}{\milli\liter}, which was filled with Ringer solution in order to experimentally replicate the electrical conductivity of the CSF ($\sigma_\text{CSF} \approx \qty{1.6}{\siemens\per\meter}$).
As the brain phantom is the only deformable element of the setup, any change in the motor displacement induces a change in the ventricular volume that is proportional to the bellows' surface area. 
The motor displacement is controlled by a cascaded proportional-integral (PI) control loop.
The outer PI controller uses the LVCM's integrated quadrature position encoder to measure the motor position $x_\text{LVCM}$ and computes a reference current $i_\text{ref}$ for the motor driver (ESCON 50/5, Maxon Motor, Sachseln, Switzerland).
The inner PI controller, which is embedded in the motor driver, tracks the motor reference current by measuring the actual motor current $i_\text{meas}$ and adjusting the pulse-width modulated input voltage $v_\text{LVCM}$ applied to the motor.
A host PC running MATLAB/Simulink Desktop Real-time (R2023a, MathWorks, Natick, MA, United States) is used for test bench operation and data acquisition (DAQ; PCIe-6321, National Instruments, Austin, TX, United States) at \qty{200}{\Hz}.

For in~vitro system validation, the measurement catheter was positioned inside the ventricle of the brain phantom as shown in Figure~\ref{fig:invitro}.
Physiological effects such as the cardiac cycle at \qty{60}{\bpm} were modeled by tracking a continuous waveform motor position reference with the LVCM.
A \qty{10}{\second} long test sequence was recorded, capturing the controlled volume along with the resulting pressure and impedance measurements.
Post-processing of the acquired data was performed by zero-phase low-pass filtering (Butterworth, 4th order, \qty{20}{\Hz} cut-off frequency).
For data analysis, the Pearson correlation coefficients between each pair of the three recorded signals were calculated.

\subsection{in~vivo Pilot Study}\label{sec:Methods_InVivo}
The in~vivo pilot study presented in this work is based on a previously established ovine model for the investigation of pathophysiological CSF dynamics~\cite{Podgorsak2022, Podgorsak2023, Trimmel2022a, Fluerenbrock2024TBME}.

\subsubsection{Ethical Statement}\label{sec:Methods_InVivo_Ethics}
Animal housing and all experimental procedures were approved by the local Committee for Experimental Animal Research (Cantonal Veterinary Office Zurich, Switzerland) under the license number ZH135/2020, and conforming to the European Directive 2010/63/EU of the European Parliament and the Council on the Protection of Animals used for Scientific Purposes, as well as to the Guide for the Care and Use of Laboratory Animals~\cite{GCULA2011}.
All health status and housing complied via formal attestation with the Swiss Federal Food Safety and Veterinary Office regulations.

\subsubsection{Experiment Preparation}\label{sec:Methods_InVivo_Setup}
One adult female Swiss White Alpine sheep (Ovis gmelini aries) was used in this pilot study.
On the day of surgery, the animal was pre-medicated intravenously with \qty[per-mode=symbol]{3}{\mg\per\kilogram} body weight (BW) ketamine hydrochloride (Ketasol-100, Dr. E. Graeub AG, Bern, Switzerland) and \qty[per-mode=symbol]{0.3}{\mg\per\kilogram} BW midazolam (Dormicum, Roche Pharma, Reinach, Switzerland). 
Prior to orotracheal intubation, anesthesia was induced with \qtyrange[per-mode=symbol]{2}{5}{\mg\per\kilogram} BW of propofol (Propofol-Lipuro \qty{1}{\percent}, B. Braun Medical AG, Sempach, Switzerland).
Anesthesia was maintained throughout the trial by inhalation of \qtyrange{1}{2}{\percent} isoflurane (Attane Isoflurane ad.us.vet., Piramal Enterprises, Lyssach, Switzerland) via volume-controlled ventilation (fresh gas flow \qtyrange[per-mode=symbol]{1}{1.5}{\liter\per\min}, \qty[per-mode=symbol]{18}{\breaths\per\min}, tidal volume \qtyrange[per-mode=symbol]{10}{15}{\milli\liter\per\kilogram} BW, FiO2~\num{0.7}, Pmax \qty{30}{\mmHg} in an oxygen/air mixture) through a semi-closed breathing circuit (Dräger Primus, Dräger Medical, Lübeck, Germany) and balanced with a constant rate infusion of propofol at \qtyrange[per-mode=repeated-symbol]{2}{4}{\milli\gram\per\hour\per\kilogram} BW. 
Intraoperative analgesia was provided by a constant rate infusion of sufentanil (Sufenta Forte, Janssen-Cilag AG, Zug, Switzerland) at \qty[per-mode=repeated-symbol]{2.5}{\micro\gram\per\hour\per\kilogram} BW. 
Ringer solution (Ringerfundin, B. Braun Medical AG, Sempach, Switzerland) was administered intravenously at an infusion rate of \qty[per-mode=repeated-symbol]{5}{\milli\liter\per\hour\per\kilogram} BW.

\begin{figure*}[!t]
    \centering
    \includegraphics[width=\textwidth]{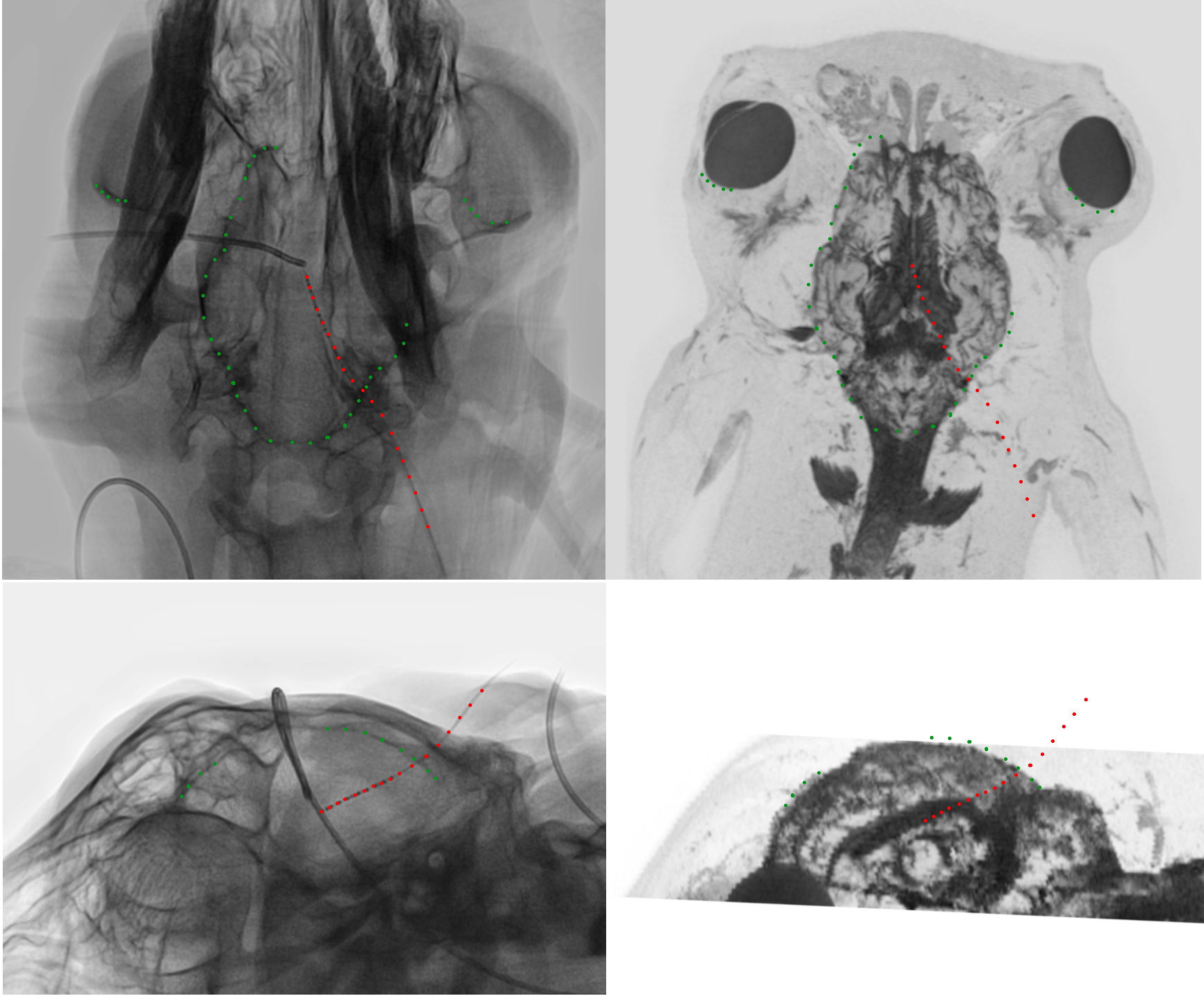} 
    \caption{Confirmation of the measurement catheter placement at the end of the in~vivo pilot study using both fluoroscopy and magnetic resonance imaging (MRI). Left: Transverse and coronal 2D fluoroscopy acquisitions of the head with implanted measurement catheter. Right: Re-sliced minimum intensity projection of the phase-sensitive Short-TI Inversion Recovery acquisition of the same brain following explantation of the measurement catheter. The green dotted lines indicate anatomical landmarks for co-registration of fluoroscopy and MRI data, whereas the red dotted lines show the trace of the measurement catheter.}
    \label{fig:implantation}
\end{figure*}

A small frontal burr hole trephination was performed using a diamond drill to access the sheep's left lateral ventricle and insert the measurement catheter for ICP and BI.
A bone wax plug (Ethicon Bone wax, Johnson \& Johnson Medical Ltd., Livingston, United Kingdom) was molded around the exiting catheter to seal the burr hole in the cranium, thus ensuring proper fixation and preventing leakage of CSF or alterations in ICP. 
A \qty{4.5}{\Fr} spinal catheter (Neuromedex, Hamburg, Germany) was placed in the intrathecal sac via a laminotomy at level L6-7 to measure intrathecal pressure (ITP).
The same access was used for the placement of a spinal needle (Perifix 310 mini set, \qty{5}{\Fr}, B.Braun Melsungen AG, Melsungen, Germany) to perform infusion experiments. 
Similar to the cranial access, the catheter and needle were again secured using bone wax.
Intravascular catheters (Avanti, Cordis Corporation, Miami Lakes, FL, United States) were placed in the carotid artery and jugular vein to measure arterial blood pressure (ABP) and central venous pressure (CVP), respectively.
Medical pressure transducers (Meritrans DTXPlus, Merit Medical Systems, Jordan, UT, United States) were used for the measurement of ABP, CVP and ITP, whereas ICP and BI were acquired with the custom-built measurement system described in Section~\ref{sec:Methods_InVitro_Design}.
All signals were sampled with a frequency of \qty{1}{\kilo\Hz} and recorded using an embedded real-time computer (MicroLabBox, dSpace GmbH, Paderborn, Germany).
The placement of the BI and ICP measurement catheter was confirmed at the end of the in~vivo pilot study using fluoroscopy and magnetic resonance imaging (MRI), as shown in Figure~\ref{fig:implantation}.

\subsubsection{Experimental Protocol}\label{sec:Methods_InVivo_Protocol}
Two data sets were recorded during the in~vivo investigation. 
First, a \qty{10}{\min} baseline measurement sequence without any kind of intervention was recorded.
Subsequently, a \qty{40}{\min} measurement sequence was recorded during which three intrathecal bolus infusions of \qty{2}{\milli\liter} Ringer solution were performed using an automated infusion apparatus~\cite{Qvarlander2013}.

\subsubsection{Data analysis}\label{sec:Methods_InVivo_Analysis}
Three types of analysis were performed for the acquired in~vivo data, all implemented in Python (version 3.13).

First, a spectral analysis of the baseline measurement data was conducted by computing the power spectral density (PSD) of each signal using Welch's method~\cite{Welch1967} with \qty{20}{\s} long Hann windows and \qty{50}{\percent} overlap.

Second, a correlation analysis was performed for both in~vivo data sets. 
The baseline measurement data was zero-phase band-pass filtered (Butterworth, 2nd order, \qty{0.5}{\Hz} and \qty{10}{\Hz} cut-off frequencies) to extract effects caused by the cardiac-induced volumetric changes and split into \qty{20}{\s} long windows. 
For each window, the normalized correlation values and corresponding lags of the maximum absolute cross-correlations were computed.
Subsequently, the median correlation values and lags across all windows were calculated.
The bolus infusion data was zero-phase low-pass filtered (Butterworth, 2nd order, \qty{0.05}{\Hz} cut-off frequency) to extract effects caused by the bolus-induced volumetric changes and split into three distinct parts corresponding to the three bolus infusions. 
For each part, Pearson correlation coefficients of relevant signal combinations were computed. 

Third, a Granger causality analysis of the bolus infusion data was performed. 
After down-sampling to \qty{1}{\Hz} and mean-centering, signal stationarity was confirmed using the Augmented Dickey–Fuller (ADF) test~\cite{Dickey1979}. 
A vector autoregressive (VAR) model was constructed with BI as the dependent variable and ABP, CVP, and ICP as predictors.
The optimal lag order for the VAR model was selected via the Akaike Information Criterion (AIC)~\cite{Akaike1974}. 
Conditional Granger causality tests~\cite{Geweke1984} were performed to evaluate whether each predictor added significant predictive information for BI after conditioning on all of the remaining signals. 
For each test, the F-statistic, p-value, and critical value at the \qty{5}{\percent} significance level are computed.

\section{Results}\label{sec:Results}
This section reports the experimental results obtained from the previously outlined in~vitro and in~vivo investigations.

\subsection{in~vitro Results}\label{sec:Results_InVitro}
The results of the in~vitro experiment are shown in Figure~\ref{fig:invitro}.
The induced changes in the ventricular volume of the brain phantom lead to significant changes in the measured BI and ICP.
Notably, the characteristic waveform pattern used to modulate the ventricular volume of the brain phantom is clearly preserved in both measurement signals. 
While changes in volume and pressure are strongly positively correlated, changes in volume and bioimpedance are strongly negatively correlated.
The exact Pearson correlation coefficients for the relationships between the three recorded signals are reported in Table~\ref{tab:results_invitro}.
\begin{table}
    \centering
    \setlength\extrarowheight{2pt}
    \caption{Pearson correlation of signals during in~vitro testing}
    \label{tab:results_invitro}
    \begin{tabularx}{\columnwidth}{XX}
    \toprule
    \textbf{Measurements} & \textbf{Correlation} \\
    \midrule
    $\Delta$ Volume - $\Delta$ ICP & 0.97 \\
    $\Delta$ ICP - $\Delta$ BI & -0.96 \\
    $\Delta$ BI - $\Delta$ Volume & -0.91 \\
    \bottomrule
    \end{tabularx}
\end{table}

\subsection{in~vivo Results}\label{sec:Results_InVivo}

\begin{table*}
    \centering
    \setlength\extrarowheight{2pt}
    \caption{Cross-correlation analysis of the in~vivo baseline data and Pearson correlation analysis of the in~vivo bolus infusion data}
    \label{tab:Results_invivo_correlation}
    \begin{tabularx}{\textwidth}{XXXXX}
    \toprule
    \textbf{Measurements} & \textbf{Baseline data} & \textbf{Bolus 1} & \textbf{Bolus 2} & \textbf{Bolus 3} \\
    \midrule
    ICP - BI & -0.71 (\qty{0.133}{\s}) & -0.19 & -0.63 & -0.22 \\
    ITP - BI & -0.72 (\qty{0.092}{\s}) & -0.18 & -0.58 & -0.20 \\
    CVP - BI & +0.68 (\qty{0.164}{\s}) & -0.30 & -0.22 & -0.20 \\
    ABP - BI & -0.72 (\qty{0.185}{\s}) & -0.05 & -0.86 & -0.71 \\
    ABP - ICP & +0.97 (\qty{0.054}{\s}) & +0.85 & +0.89 & +0.82 \\
    ABP - ITP & +0.90 (\qty{0.096}{\s}) & +0.84 & +0.87 & +0.81 \\
    ABP - CVP & -0.82 (\qty{0.030}{\s}) & +0.81 & +0.64 & +0.80 \\
    ICP - ITP & +0.85 (\qty{0.042}{\s}) & +1.00 & +0.99 & +1.00 \\
    \bottomrule
    \end{tabularx}
\end{table*}
  
\begin{table*}
    \centering
    \setlength\extrarowheight{2pt}
    \caption{Granger causality analysis of the in~vivo bolus infusion data}
    \label{tab:Results_invivo_causality}
    \begin{tabularx}{\textwidth}{XXXXX}
    \toprule
    \textbf{Prediction} & \textbf{Conditioned on} & \textbf{p-value} & \textbf{F-statistic} & \textbf{Critical value} \\
    \midrule
    ICP $\rightarrow$ BI & ABP, CVP & $<$0.001 & 29.028 & 1.832 \\
    ABP $\rightarrow$ BI & ICP, CVP & $<$0.001 & 10.773 & 1.832 \\
    CVP $\rightarrow$ BI & ICP, ABP & $<$0.001 & 6.158 & 1.832 \\
    \bottomrule
    \end{tabularx}
\end{table*}

\begin{figure*}
    \centering
    \includegraphics[width=\textwidth]{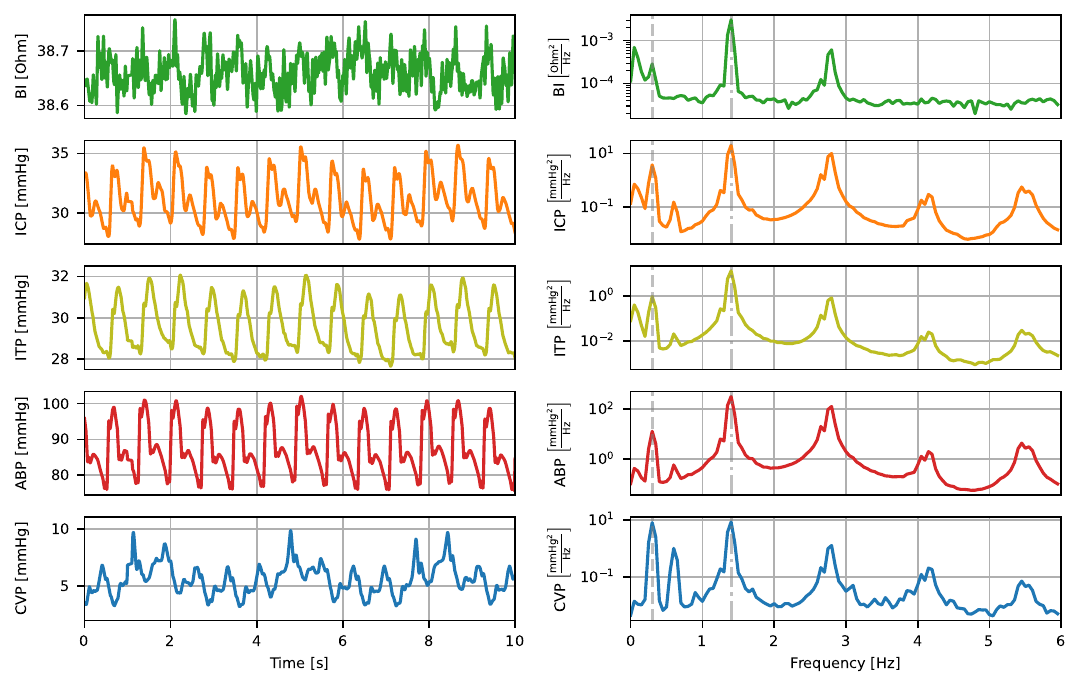} 
    \caption{Left: Exemplary time series data extracts of the acquired bioimpedance (BI), intracranial pressure (ICP), intrathecal pressure (ITP), arterial blood pressure (ABP), and central venous pressure (CVP) during the in~vivo baseline measurement. Right: Power spectral densities of the respective signals computed from the full baseline measurement using Welch's method (logarithmic y-axis). Dashed lines at \qty{0.3}{\Hz} indicate the respiration rate, whereas dashed-dotted lines at \qty{1.4}{\Hz} indicate the heart rate.}
    \label{fig:baseline}
\end{figure*}

\begin{figure*}[!t]
    \centering
    \includegraphics[width=\textwidth]{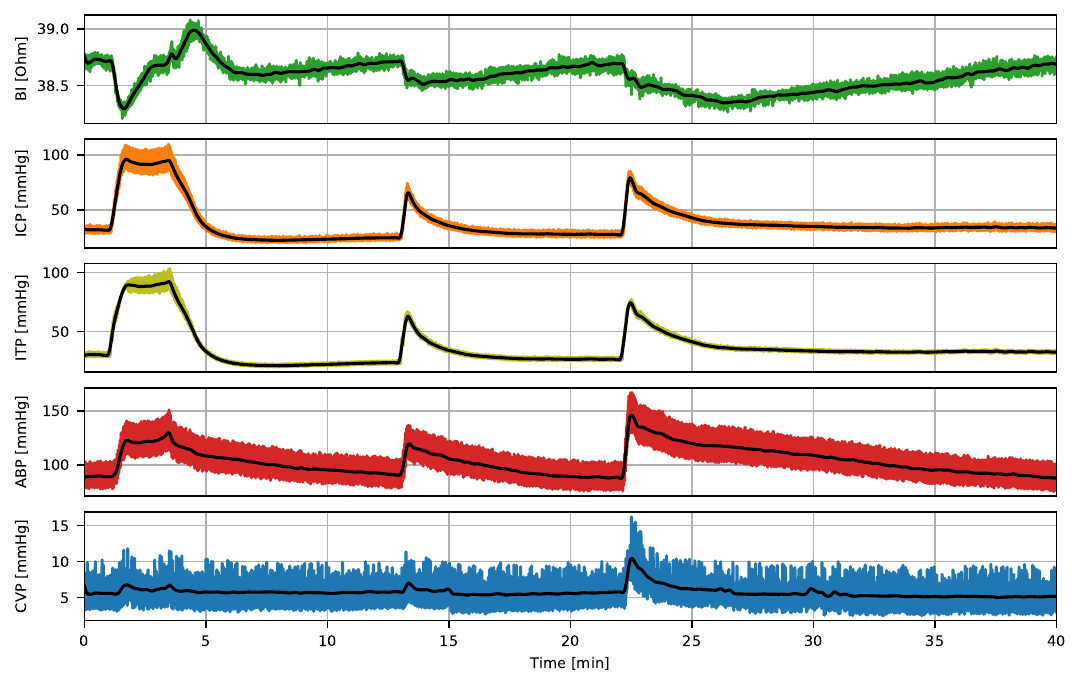} 
    \caption{Time series data of the acquired bioimpedance (BI), intracranial pressure (ICP), intrathecal pressure (ITP), arterial blood pressure (ABP), and central venous pressure (CVP) during the in~vivo experiment with intrathecal bolus infusions of Ringer solution. The black lines show the baseline components of the respective signals computed via zero-phase low-pass filtering with \qty{0.05}{\Hz} cut-off frequency.}
    \label{fig:bolus}
\end{figure*}

A representative extract of the acquired time-series data during the in~vivo baseline measurement is shown in Figure~\ref{fig:baseline}. 
ICP, ITP, ABP, and CVP all exhibit characteristic physiological waveforms reflecting both cardiac and respiratory influences.
BI shows similar waveform components, albeit with a higher level of measurement noise. 
The corresponding power spectral densities of the measured signals are also presented in Figure~\ref{fig:baseline}. 
In all signals, respiration-related peaks were consistently observed around \qty{0.3}{\Hz} and cardiac-related peaks around \qty{1.4}{\Hz}. 

The signal responses during the bolus infusion experiments are illustrated in Figure~\ref{fig:bolus}.
Following each of the three intrathecal bolus infusion of Ringer solution, ICP, ITP, and ABP exhibit pronounced increases in pressure, whereas CVP shows only minor fluctuations.
BI decreases consistently after each of the three bolus infusions, reflecting sensitivity to the active alteration of the CSF volume.
While bolus infusions 2 and 3 show clear and comparable dynamics across all recorded signals, the response to bolus infusion 1 differs. 
Here, a rapid rise in ITP exceeded the safety threshold of the automated infusion apparatus~\cite{Qvarlander2013}, causing an intermittent infusion delivery that in turn prolonged the bolus infusion.
Compared to bolus infusions 2 and 3, this irregular infusion pattern resulted in altered temporal dynamics in the BI signal as well as the contextual CSF and blood pressure signals.

The results of both correlation analyses are summarized in Table~\ref{tab:Results_invivo_correlation}. 
During baseline conditions, BI exhibited strong negative correlations with ABP and CSF pressures.
Median cross-correlation values exceeding $-0.7$ were accompanied by relatively short lag times \qtyrange{0.092}{0.185}{\s}.
Pearson correlation values during bolus infusions were lower in magnitude, yet ABP and ICP remained the most strongly correlated signals of BI. 
Strong correlations were also observed between ABP, ICP, and ITP, reflecting their well-established strong physiological coupling. 

The results of the conditional Granger causality analysis are presented in Table~\ref{tab:Results_invivo_causality}.
All predictors were found to provide significant additional information for predicting BI even when conditioned on the remaining predictor signals. 
Among the three predictors, ICP showed the strongest causal contribution, followed by ABP and CVP.

\section{Discussion}\label{sec:Discussion}
The presented results of this work are consistent with the findings from previous in~vitro~\cite{Linninger2009a, Castelar2019} and in~vivo~\cite{Basati2011, Basati2012} studies, providing further evidence for the hypothesized relationship between BI and CSF volume.
While the previous in~vivo studies relied on small animal models, the use of a large animal model in the presented pilot study additionally allowed for concurrent measurements of physiologically related blood and CSF pressures.
The spectral analysis of the baseline data indicates that characteristic waveform features in the BI signal reflect underlying physiological activity. 
Furthermore, the complementary correlation and Granger causality analyses of the bolus infusion data show that both CSF and blood pressures contribute independently to modulations in the BI signal.

While large animal models enable the replication of complex pathophysiological processes and the testing of medical devices under realistic conditions that facilitate the knowledge transfer to applications in humans~\cite{McAllister2015, Armstead2020}, the primary objective of this study was to provide a proof-of-principle for the developed bioimpedance measurement system.
Adherence to the 3R principles (Replacement, Reduction, Refinement)~\cite{Russell1959, Hubrecht2019} limited this pilot therefore to a single subject and the presented results remain illustrative rather than statistically conclusive.

Nonetheless, two key considerations for future research emerge from this study.
First, cerebral hemodynamics and blood flow should be accounted for when modeling and analyzing intraventricular bioimpedance. 
While previous studies mainly focused on the brain tissue and CSF~\cite{Linninger2009a, Castelar2019, Basati2011, Basati2012}, ABP exhibits one of the strongest correlations with BI among all measured signals in this study.
Although ABP is considered as one of the main drivers of ICP~\cite{Linninger2016, Ohnemus2025} and could thereby indirectly influence BI, the Granger causality analysis indicates that both ABP and CVP influence BI even after conditioning on ICP, suggesting a direct contribution.
Alterations in the cerebral blood flow can occur due to cardiac pulsations and autoregulatory mechanisms~\cite{Cushing1901, Schmidt2018}, as evident in the ABP response following bolus infusion of artificial CSF in Figure~\ref{fig:bolus}. 
Given that blood has a lower electrical conductivity than CSF (see Table~\ref{tab:electrical_conductivity}), these cerebral blood flow alterations may significantly affect BI measurements by changing the ratio of brain tissue, CSF and blood within the measurement field.
Second, the exact positioning and potential movement of the measurement catheter, e.g., due to shifts in the cranial tissue distribution resulting from the cardiac cycle or bolus infusions of artificial CSF, should be investigated using medical imaging technologies such as continuous MRI in combination with MRI-conditional measurement catheters.
Since the spatial sensitivity of the electromagnetic field and its influence on the impedance measurement has already been demonstrated in the application of conductance catheters for the measurement of ventricular heart volume~\cite{Larson2013, Korn2021}, it can be expected to be similarly relevant in the application of conductance catheters for the measurement of ventricular CSF volume.

\section{Conclusion}\label{sec:Conclusion}
This work presented the development and application of a measurement system for intraventricular BI. 
Following in~vitro validation using a mechatronic test bench, the measurement system was used in an in~vivo pilot study with contextual CSF and blood pressure measurements.
The results of the time series analysis not only revealed physiological BI waveform components linked to cardiac and respiratory cycles but also significant changes in BI following intrathecal bolus infusions of artificial CSF.
These findings highlight the potential of BI as an advanced measurement modality for hydrocephalus patient monitoring. 
Smart shunt systems for hydrocephalus therapy~\cite{Misgeld2015, Fluerenbrock2025FBCNS} could augment measurements of ICP and CSF drainage with BI to facilitate CSF volume monitoring~\cite{Fluerenbrock2023AUTOMED, Fluerenbrock2024CCTA} and physiologically informed CSF drainage control.
To realize this vision, however, further research is required to fully understand the influence of cerebral hemodynamics and catheter placement on BI measurements to establish its reliability across subjects and pathological conditions.

\section*{Acknowledgements}
This work was supported by the Swiss National Science Foundation (SNSF) through grant 315230 184913.

\bibliographystyle{elsarticle-num} 
\bibliography{references.bib}
\end{document}